\documentclass{aa}  

\usepackage{graphicx}
\usepackage{txfonts}
\usepackage[colorlinks=true, citecolor=blue, linkcolor=blue]{hyperref}
\usepackage{xcolor}
\usepackage[normalem]{ulem}
\usepackage{natbib}
\bibpunct{(}{)}{;}{a}{}{,} % to follow the A&A style

\makeatletter
\renewcommand*\aa@pageof{, page \thepage{} of \pageref*{LastPage}}
\makeatother
\begin{document}

   \title{Apparent motion of penumbral grains in a sunspot simulation}

   \subtitle{}

   \author{Michal Sobotka \inst{1}
          \and
          Markus Schmassmann \inst{2}
          }

   \institute{Astronomical Institute of the Czech Academy of Sciences,
              Fri\v{c}ova 298, 25165 Ond\v{r}ejov, Czech Republic \\
              e-mail: {\tt michal.sobotka@asu.cas.cz}
   \and
    Institute for Solar Physics (KIS), Georges-K\"ohler-Allee 401a, 79110 Freiburg, Germany
 }

  \date{Received 4 October 2024; accepted 7 June 2025}

\abstract
  % context heading (optional) {} leave it empty if necessary
   {The bright heads of penumbral filaments, penumbral grains (PGs), are manifestations of hot plasma flows rising to the surface. They are observed to move horizontally toward the sunspot umbra or away from it. Recent analyses of observations indicate that the direction of this motion is related to the inclination of the surrounding magnetic field.}
   {The penumbra of a sunspot simulated by the radiative magnetohydrodynamic code MURaM is analysed to get typical physical conditions in PGs, compare them to those in the surroundings, describe their spatial distribution, and study their evolution.}
   {We use time series of images that map intensity, temperature, magnetic field vector, and velocity vector in horizontal slices at the visible surface, in subsurface layers, and in vertical cuts through the simulation box to track PGs and compare, statistically and in individual cases, the physical quantities inside them with those in the surroundings.}
   {The statistical analysis of simulation results provides average values of temperature, magnetic field strength and inclination, vertical velocity, and their changes with radial distance from the spot centre. We find a subtle difference between simulated PGs with opposite directions of motions when comparing the magnetic field inclinations inside and outside the PGs. The case studies, documented by movies, show that the differences of inclinations and the direction of motions may change during the lifetime of some PGs and that the turbulence in the surface layers introduces some randomness in the apparent motions of PGs.}
   {}

  \keywords{sunspots -- Sun: photosphere -- magnetohydrodynamics (MHD)}

  \titlerunning{Apparent motion of simulated penumbral grains}

\maketitle
\nolinenumbers   % removes line numbering
%
%%%%%%%%%%%%%%%%%%%%%%%%%%%%%%%%%%%%%%%%%%%%%%%%%%%%%%%%%%%%%%%%%%%%%%%%

\section{Introduction}
\label{sec:intro}

A sunspot penumbra shows a radial filamentary structure in white-light images. This appearance has been plausibly explained by \citet{Tiwari2013}, who presented semi-empirical models of penumbral filaments derived from inversions of spectropolarimetric observations. Hot plasma from subphotospheric layers enters a filament at its inner end, the filament's head, oriented towards the umbra \citep{Ichimoto2007, Franz2009}. The plasma flow becomes horizontal in the filament's body, constituting the Evershed flow \citep{Evershed1909}, and also leaks laterally, producing downflows on the sides of the filament \citep[cf.][]{Scharmer2011Sci, Joshi2011ApJ}. The plasma cools down by radiation so that the filament's body gets cooler and darker with the increasing distance from the filament's head. Finally, the cool plasma turns down at the end of the filament \citep{Tiwari2013}.
This flow pattern is embedded in a background magnetic field, spines \citep{Lites1993}, the inclination of which gradually increases with the distance from the sunspot umbra. On the other hand, the magnetic field in the bodies of filaments is nearly horizontal \citep[e.g.,][]{SolankiMont1993, Tiwari2013}. The filaments represent narrow and long magnetoconvective cells, which are present everywhere in the penumbra and are bright near their heads and dark near their ends. This model is consistent with the magnetohydrodynamic (MHD) numerical simulations of sunspot structure \citep{Rempel2012}.

The heads of filaments are called penumbral grains \citep[PGs,][]{Muller1973}. They are observed to move toward the umbra (inwards: INW) and also away from the umbra \citep[outwards: OUT,][]{SBSIII1999, SSIV2001, Zhang2013,SobPusch2022}. The aforementioned works stated that the INW PGs are concentrated in the inner penumbra, while the OUT PGs mostly appear in the outer penumbra and, in total, are less numerous than the INW PGs. The motion of PGs is apparent and does not correspond to the real motion of the plasma, namely, an upflow and horizontal outflow. 

The inward motion of PGs can be explained by a buoyancy force that makes an initially inclined column of upflowing hot plasma more vertical, such that the intersection of the column with the visible surface (PG) moves radially toward the umbra \citep{Schlich1998AA}. A possible hint to explain the outward motion can be found in the semi-empirical models of \citet{Tiwari2013}. At the optical depth $\tau_{500\rm nm} = 1$, the magnetic field inclination in PGs is larger (more horizontal) than the surrounding one in the inner penumbra and smaller (more vertical) in the outer penumbra. Because the INW PGs dominate in the inner penumbra and the OUT PGs appear in the outer one, the opposite directions of motions may be related to the differences between the magnetic inclinations inside PGs and in the surrounding magnetic field of spines \citep{SobPusch2022}: Rising hot plasma in a PG surrounded by a less inclined magnetic field may adapt its trajectory to be more vertical, resulting in an inward apparent motion of the PG. Conversely, it may be dragged by a more horizontal surrounding field such that an outward apparent motion is observed.

\citet{Sobotka2024} tried to verify this scenario, using height-stratified inversions of spectropolarimetric observations of five sunspot penumbrae with a spatial resolution better than 0\farcs 16 per pixel, obtained with the Hinode satellite \citep{Kosugi2007Hino}, the Swedish Solar Telescope \citep{SchaSST2003}, and the GREGOR solar telescope \citep{Schmidt2012}. On a sample of 444 INW PGs and 269 OUT PGs, \citet{Sobotka2024} showed that 43\% of the INW PGs had their magnetic inclination more horizontal than the inclination in their surroundings and 50\% of the OUT PGs had the inclination more vertical than the surrounding one. The opposite differences of inclinations were found in only 22\% of the INW and 19\% of OUT PGs. In the rest of the cases (35\% of INW and 31\% of OUT PGs), the determination of the inclination differences was not reliable, which was explained by limitations in spatial resolution and accuracy of spectropolarimetric inversions. This result confirmed a certain statistical relation between the direction of motions of the PGs and the difference between magnetic field inclinations inside and outside the PGs.
In this paper, we analyse physical conditions in and around PGs that appear in numerical MHD simulations of a sunspot.

\citet{2009Sci...325..171R} %Rempel
introduced the first realistic radiative MHD simulations of round sunspots. This simulation was continued at a higher resolution, and the penumbral fine structure was discussed in \citet{Rempel2011}. %2011ApJ...729....5R
These simulations were further improved into those of \cite{Rempel2012}, %2012ApJ...750...62R
by forcing the field at the top boundary to be more inclined than for a potential field 
(\citealp[see][appendix B]{Rempel2012}; %2012ApJ...750...62R
\citealp[][appendix C.1]{2006PhDT.......338C}). %Cheung
The simulation in \citet{2015ApJ...814..125R} %Rempel
is similar, with the advantage of a larger simulation box, but at a lower resolution than is available for \cite{Rempel2012}. %2012ApJ...750...62R

\citet{2020A&A...638A..28J} %Jurčák
showed that the magnetic field distributions of these types of non-potential simulations are inconsistent with observations, in particular, at the umbral boundary the vertical component of the magnetic field of the simulations is too weak and the average field inclination is too high (too close to horizontal). Within the penumbra itself, the horizontal field is too strong. Furthermore, the umbra is too large compared to the spot. 
While other simulations based on a potential-field initial condition \citep[e.g.][and in prep.]{2024eas..conf.1941S} show less of a discrepancy in the magnetic field distributions in the penumbra, for example, Type I in \citet{2020A&A...638A..28J}, %Jurčák
they fail to create the Evershed flow along the full length of the penumbral filaments, and the upflows are not concentrated at the inner end of these filaments.
These potential simulations do not show clear PGs, making them unsuitable for our analysis here.

In this article, Sect.~\ref{sec:numer} introduces the numerical sunspot model, Sect.~\ref{sec:analysis} describes the data analysis, Sect.~\ref{sec:stat} the results of the statistical analysis, Sect.~\ref{sec:indiv} the study of individual cases, and Sect.~\ref{sec:discuss} wraps up with discussion and conclusions.

%%%%%%%%%%%%%%%%%%%%%%%%%%%%%%%%%%%%%%%%%%%%%%%%%%%%%%%%%%%%%%%%%%%%%%%%

\section{Numerical model of a sunspot}
\label{sec:numer}

The sunspot simulation was carried out with the radiative MHD code MURaM \citep{Vogler:2005, 2009ApJ...691..640R}. %2005A&A...429..335V, Rempel
It used non-gray radiative transfer in a box of $49.152^2\times6.144\,\mathrm{Mm}^3$ with grid cell sizes of 32~(16)\,km horizontally (vertically). It is a continuation of a run described in \cite{Rempel2012} %2012ApJ...750...62R
as $\alpha=2$. A short description of the initial field is in Appendix~\ref{sec:app_ini_mag}. The upper boundary forces the horizontal component of the field to be twice as large as it would be for a potential field, as described in equations~B3 of \citet{Rempel2012}, %2012ApJ...750...62R, 
resulting in a more inclined field. This increases the width of the penumbra to half of the spot radius.
Until 05:24:47 of the simulation time its grid cells only had a resolution of 48~(24)\, km. Then the simulation was continued at the higher resolution of 32~(16)\,km. In this form, it was described in \cite{2020A&A...638A..28J}. %Jurcak
The current work uses data from 05:54:47-06:24:47, which differs from the previous half-hour only by a new Equation of State (EOS) table and updated MURaM version \citep[Sect.~1.5 of][]{2017ApJ...834...10R}. %Rempel
From the EOS table, pressures and temperatures are retrieved given densities and internal energies per mass.
The new EOS table differs only in the chromosphere, and the only change in the MURaM code affecting the data for this article is that $I_\mathrm{out}$ was calculated at 500\,nm instead of using the first opacity bin, and 500\,nm opacities were used to calculate the height of the $\tau=1$ layer. This makes the umbra appear darker in the $I_\mathrm{out}$ image of the newer simulations. The first opacity bin contains light from wavelengths, which have $\tau_\lambda=1$ deepest in the atmosphere \citep[see][]{2004A&A...421..741V}, %Vögler
thereby corresponding to the continuum; it contains on average $\approx90\%$ of the radiative energy.

The data used for the analysis in this article are intensity, $I_\mathrm{out}$, images calculated using 500\,nm continuum opacities, the magnetic field vectors, velocity vectors, gas pressure and temperatures at the \mbox{$\tau_{500\rm nm}=1$} iso-surface (hereafter slices) and vertical cuts through the centre of the spot. We also utilise two sequences of the above-mentioned variables in subphotospheric slices at the depths of 160~km (10 grid steps) and 320~km (20 grid steps) below the $\tau_{500\rm nm}=1$ iso-surface, as well as the 3D box at the middle of our simulation run. The time sequences of 101 intensity images, surface variables, and vertical cuts are available at a cadence of $\approx18$\,s and the sequences of variables in subphotospheric slices have 11~elements with the cadence of $\approx180$\,s. The axes are oriented such that vertical velocity components $v_{\rm z}>0$ are upflows. Additionally, we calculate plasma beta, the ratio of gas and magnetic pressures $\beta = 8\pi P_g/B^2$, in each slice and vertical cut.

The gas in the low-density region above the sunspot does not have enough inertia to influence the behaviour at and below the surface significantly. This allows for manipulations of the thermodynamics to keep sound speeds low enough and time steps short. While some form of manipulation is found in many sunspot simulations, in the one presented here we set the internal energy per mass to a fixed value in regions where the Alfv{\'e}n speed limiter \citep{Boris1970,2017ApJ...834...10R} is active. A side-effect of this can be stripes and chessboard patterns visible in vertical cuts of velocity well above the photosphere.

The umbral fraction of our simulation is too high. The increase of the inclination with increasing radial fraction is consistent with observations. This results in the inner part of the penumbra of the simulation having a larger average field inclination than in observed sunspots. Furthermore, the field strengths in the penumbra are generally too high.

%%%%%%%%%%%%%%%%%%%%%%%%%%%%%%%%%%%%%%%%%%%%%%%%%%%%%%%%%%%%%%%%%%%%%%%%

\section{Data analysis}
\label{sec:analysis}

We analyse the simulations in a similar way as the observations presented in \citet{Sobotka2024} and extend it to include a comparison of physical quantities at different distances from the centre of the sunspot and at different depths. Additionally, we inspect vertical plane cuts across PGs to get information about the height stratifications of simulated quantities and their evolution in time. The horizontal dimensions of the simulation results are reduced to 34.528 $\times$ 34.528~Mm$^2$ including only the sunspot.

The time sequence of 101 slices of intensity $I_{\rm out}$, vertical velocity component $v_{\rm z}$, magnetic inclination $\gamma$, magnetic field strength $B$, and temperature $T$ at the surface corresponding to the optical depth $\tau_{500\rm nm} = 1$ produced by the simulation run was used for the statistical analysis of $\gamma$, $B$, $T$, and $v_{\rm z}$ in and around PGs. The time step of the sequence was 18 s. The sunspot area was delimited by the threshold of $0.87\, I_{\rm ph}$ in the $I_{\rm out}$ map smoothed by a boxcar of $1.7 \times 1.7$~Mm, where $I_{\rm ph} = 1$ is the mean intensity of undisturbed granulation. PGs were defined by the thresholds $I_{\rm out} > 0.45\, I_{\rm ph}$ and $v_{\rm z} > 1.5$~km~s$^{-1}$. These values were obtained empirically after several trials to select bright and well-marked PGs and to exclude umbral structures. A sequence of segmentation masks with PGs that met the criteria mentioned above was made from the $I_{\rm out}$ and $v_{\rm z}$ series of slices.

\begin{figure}[t]\centering
\includegraphics[width=0.49\textwidth]{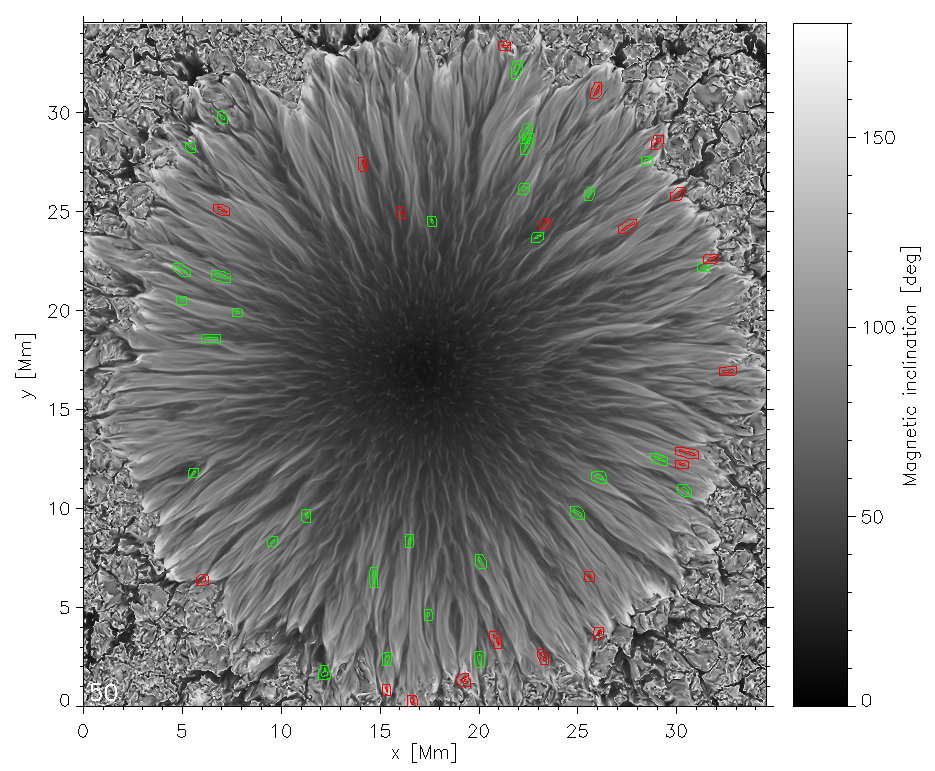}
\caption{Magnetic-field inclination slice No.\,50 at $\tau_{500\rm nm} = 1$ together with contours of identified PGs and their surroundings (green -- inward-moving; red -- outward-moving). A movie is available {\it online}. % Fig1movie.mp4
\label{fig:inc53}}
\end{figure}

The segmented $I_{\rm out}$ slices served as the input for a feature tracking pipeline based on the code written by \citet{SBSI1997} and modified by \citet{Hamedivafa2008}. The procedure identified and recorded intensities, positions, areas, and lifetimes in the samples of INW and OUT PGs that did not pass any split or merge event with other objects to avoid overlapping of filaments, lived longer than 3~min, their minimum areas were 9 cells (9216~km$^2$), and their intensity maxima were displaced more than 160~km during their lifetimes. The magnetic inclinations in and around these PGs at $\tau_{500\rm nm} = 1$ were obtained using the corresponding sequence of $\gamma$ slices. The mean $\gamma_{\rm PG}$ inside each PG was averaged over the area of the PG in the segmentation mask. The surrounding area was obtained by applying the dilate operator to the PG area, extending it by 384~km (0\farcs 53) around its perimeter, and subtracting the PG area from the result. The mean surrounding inclination $\gamma_{\rm s}$ was calculated in that area. The amount of extension was selected to maximise the differences between $\gamma_{\rm PG}$ and $\gamma_{\rm s}$ and it was consistent with an analogous parameter $d$ used by \citet{Sobotka2024} for observational data. Mean values of other physical quantities ($B$, $T$, $v_z$) were obtained likewise. An example of a magnetic-inclination slice together with contours of identified PGs (green -- INW, red -- OUT) and their surroundings is shown in Fig.~\ref{fig:inc53} and the complete movie is available {\it online}. % ...movie_01.mp4
The areas of PGs and their surroundings were also projected down to two slices of subphotospheric layers, 160~km and 320~km below the surface, to obtain corresponding values of physical quantities.

\begin{figure}[t]\centering
\includegraphics[width=0.413\textwidth]{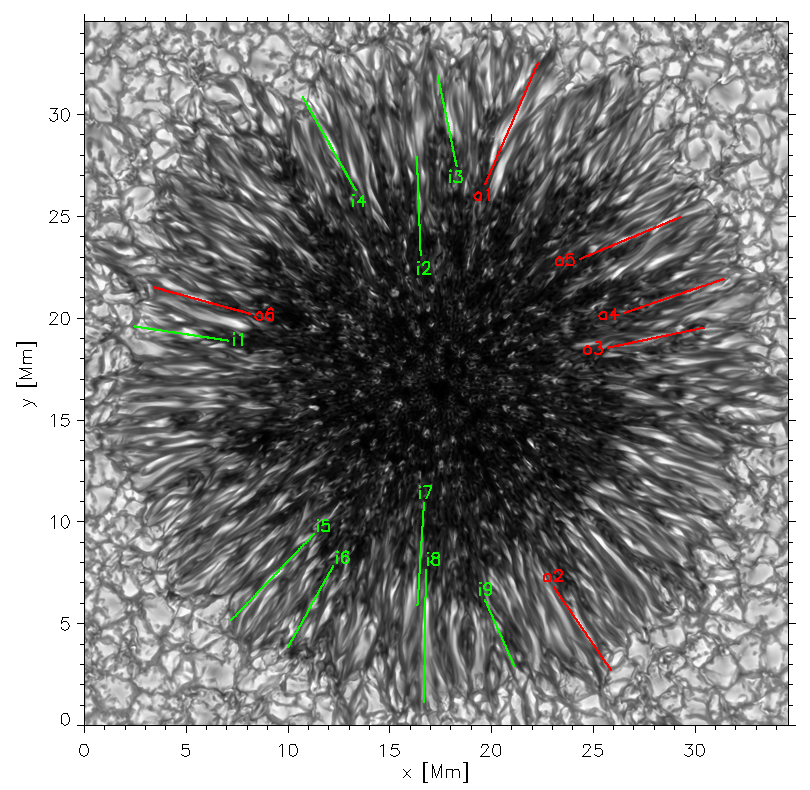}
\caption{Intensity slice No.\,50 together with the positions of static vertical cuts {\it i1--i9} (green) and {\it o1--o6} (red) across the inward-moving and outward-moving PGs, respectively.
\label{fig:cutmap}}
\end{figure}

The vertical plane cuts through PGs are important for understanding the physical processes in their vicinity. For the reference height $z = 0$, approximating the visible surface, we adopted the median height of the $\tau_{500\rm nm} = 1$ iso-surface in the penumbral area.
We visually selected nine cuts through INW and six cuts through OUT PGs with distinct apparent motions in the snapshot of the simulation box (time step No.\,50), using a short movie composed of $I_{\rm out}$ slices No.\,45--55 (3 min). The positions of corresponding cuts are marked in the $I_{\rm out}$ slice No.\,50 shown in Fig.~\ref{fig:cutmap} (green -- INW, red -- OUT). Shortenings of the cuts due to projections to the $x$ or $y$ axis were corrected to obtain the actual lengths. The range of heights $z$ was from $-1280$~km below the approximate visible surface to $+480$~km above it (111 grid steps in total). The resulting vertical cuts in temperature $T$, magnetic field strength $B$, magnetic inclination $\gamma$, vertical velocity $v_{\rm z}$, and horizontal velocity $v_{\rm h}=\sqrt{\smash[b]{v_{\rm x}^2+v_{\rm y}^2}}$ are described in Sect.~\ref{sec:indiv}.

From the time sequence of 101 vertical cuts in the $xz$ plane crossing the sunspot centre, we selected visually seven INW and four OUT PGs that coincided with the cutting plane. The range of heights $z$ was $-1120$ to 480~km (101 grid steps). The selected PGs appear in ten {\it movies} % vmv01-10.zip
that display $T$, $B$, $\gamma$, and $v_{\rm z}$ with the time step of 18~s and are discussed in Sect.~\ref{sec:indiv}.

%%%%%%%%%%%%%%%%%%%%%%%%%%%%%%%%%%%%%%%%%%%%%%%%%%%%%%%%%%%%%%%%%%%%%%%%

\section{Results}
\label{sec:results}

We present statistics of magnetic inclinations $\gamma$, magnitudes of magnetic field strengths $B$, temperatures $T$, and vertical components of velocities $v_z$ inside and outside PGs detected automatically in the sequence of slices at the visible surface and in the sequences of slices at 160~km  and 320~km below the surface.
We get more details in individual studies of PGs in different vertical cuts and in the time sequence of $xz$ cuts crossing the sunspot centre.

\subsection{Statistical analysis}
\label{sec:stat}

From the sequence of 101 $I_{\rm out}$, $v_{\rm z}$, $\gamma$, $T$, and $B$ slices at $\tau_{500\rm nm} = 1$, processed by the feature-tracking pipeline, we obtained the sample of 226~INW and 107~OUT PGs; 189 PGs did not show apparent motions larger than 160~km during their lifetime and they were excluded from the statistics. For each of INW and OUT PGs we calculated mean values of inclinations $\gamma_{\rm PG}$ inside PG and $\gamma_{\rm s}$ in its surroundings from the sequence of $\gamma$ slices and similarly all other physical quantities from the sequences of corresponding slices (Sect.~\ref{sec:analysis}).

We compare the inclinations for all the occurrences of PGs in individual frames using the following classification \citep[cf.][]{Sobotka2024}. Initial class $-1$: $\gamma_{\rm PG} < \gamma_{\rm s}$, that is, the magnetic field is more vertical in the PG than in the surroundings. Initial class 1: $\gamma_{\rm PG} > \gamma_{\rm s}$, that is, the magnetic field is more horizontal in the PG than in the surroundings.
The INW and OUT PGs have 4163 and 1781 occurrences in individual frames, respectively. Of the INW PGs, 53\% are assigned initial class 1 and 47\% class $-1$. This means that the case of $\gamma_{\rm PG} > \gamma_{\rm s}$ is slightly preferred. The small difference between these populations of classes is caused by an absence of numerous INW PGs in the inner penumbra, which commonly appear in observations but are rarely produced by the simulation (cf. Sect.~\ref{sec:numer} and Fig.~\ref{fig:cutmap}). Of the OUT PGs, 37\% have initial class 1 and 63\% class $-1$, therefore the case of $\gamma_{\rm PG} < \gamma_{\rm s}$ prevails here.

\begin{figure}[t!]\centering
\includegraphics[width=0.49\textwidth]{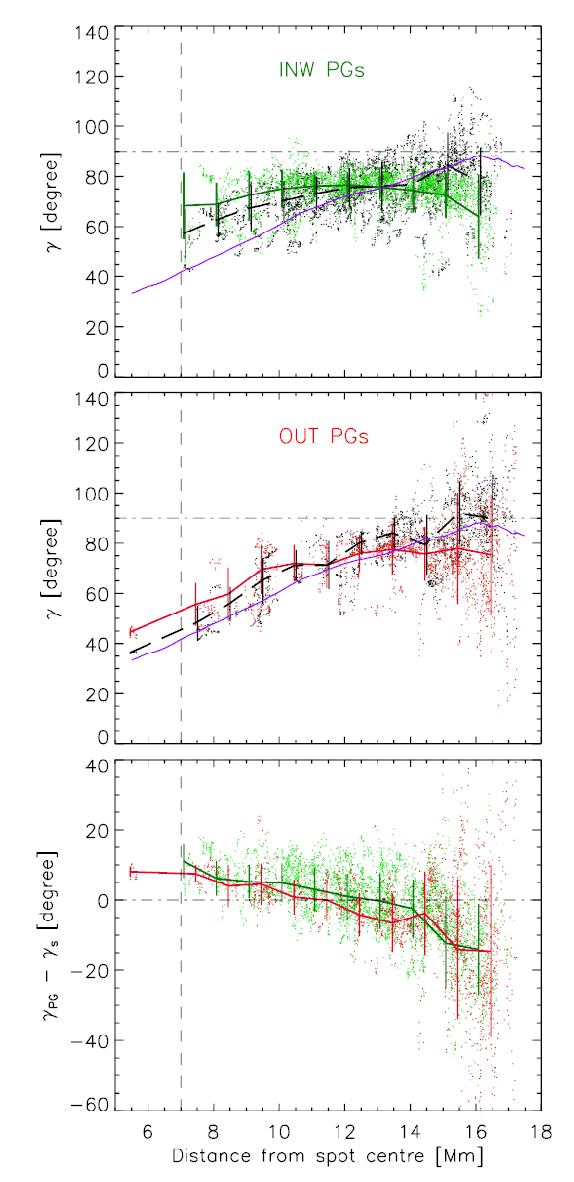}   % .eps better
\caption{Scatter plots of magnetic inclinations $\gamma$ versus radial distance from the spot centre in individual occurrences of PGs (green -- inward-moving, top panel; red -- outward-moving, middle panel) and in their surroundings (black) at $\tau_{500\rm nm} = 1$. Lines with error bars were obtained by averaging the values in 1 Mm wide sections of the radial distance. Magenta lines represent the azimuthally averaged magnetic field inclination. The vertical dashed lines mark the umbra/penumbra border usual for observed circular sunspots ($0.4\, r_{\rm spot}$). The bottom panel shows differences between the inclinations in INW and OUT PGs and their surroundings, their averages, and standard deviations.
\label{fig:inc_rdist}}
\end{figure}

Magnetic field inclinations of INW (green), OUT (red) PGs, and their surroundings (black) at different radial distances, $r$, from the spot centre, together with azimuthally averaged inclination (magenta line) at $\tau_{500\rm nm} = 1$, are plotted in Fig.~\ref{fig:inc_rdist}. The scatter of individual values is large. We show average values and standard deviations of inclinations (lines with error bars) calculated in 1~Mm wide sections of the radial distance from the spot centre.
We can see that the mean inclinations of INW PGs are approximately constant with the radial distance, ranging between $70^\circ$ and $77^\circ$. The mean inclinations of OUT PGs behave similarly in the middle and outer penumbra but they are lower ($45^\circ$--$60^\circ$) in the inner penumbra. The mean inclinations in the surroundings around OUT PGs follow the trend of the azimuthally averaged inclination, while those of INW PGs are higher than that in the inner and middle penumbra. The inclinations obtained in these simulations are higher than the observed ones by about $15^\circ$ \citep[e.g.,][]{Tiwari2013}, which is given by the simulation setting (Sect.~\ref{sec:numer}).
The azimuthally averaged inclination increases monotonously from $40^\circ$ at the umbra-penumbra boundary to nearly $90^\circ$ at the outer penumbral border. It is lower than the mean inclination in the surroundings of PGs in the inner and middle penumbra, because in addition to penumbral filaments it also includes low-inclined spines.

Both for INW and OUT PGs, the average differences $\gamma_{\rm PG} - \gamma_{\rm s}$ are positive, approximately \mbox{$5^\circ$--$10^\circ$} in the inner penumbra \mbox{($r < 10$\,Mm)}, they approach to zero in the middle penumbra ($10 < r < 13$\,Mm) and become negative, $-5^\circ$-- $-15^\circ$, in the outer penumbra ($r > 13$\,Mm). The change of sign occurs first for the OUT PGs at the distance of 11~Mm and then for the INW PGs at 13~Mm, as seen in the bottom panel of Fig.~\ref{fig:inc_rdist}. On average, INW PGs have the initial class~1 in the inner and middle penumbra and class~$-1$ in the outer one, while OUT PGs have a small population of class 1 in the inner and large population of class $-1$ in the middle and outer penumbra. We can see that $\gamma_{\rm PG} - \gamma_{\rm s}$ behave differently for INW and OUT PGs but, taking into account the values of standard deviations around $6^\circ$--$10^\circ$ and up to $20^\circ$ near the outer edge of the penumbra, these differences are quite subtle, comparable with the scatter of individual values. 

\begin{figure*}[t]\centering
  \sidecaption
  \includegraphics[width=12cm]{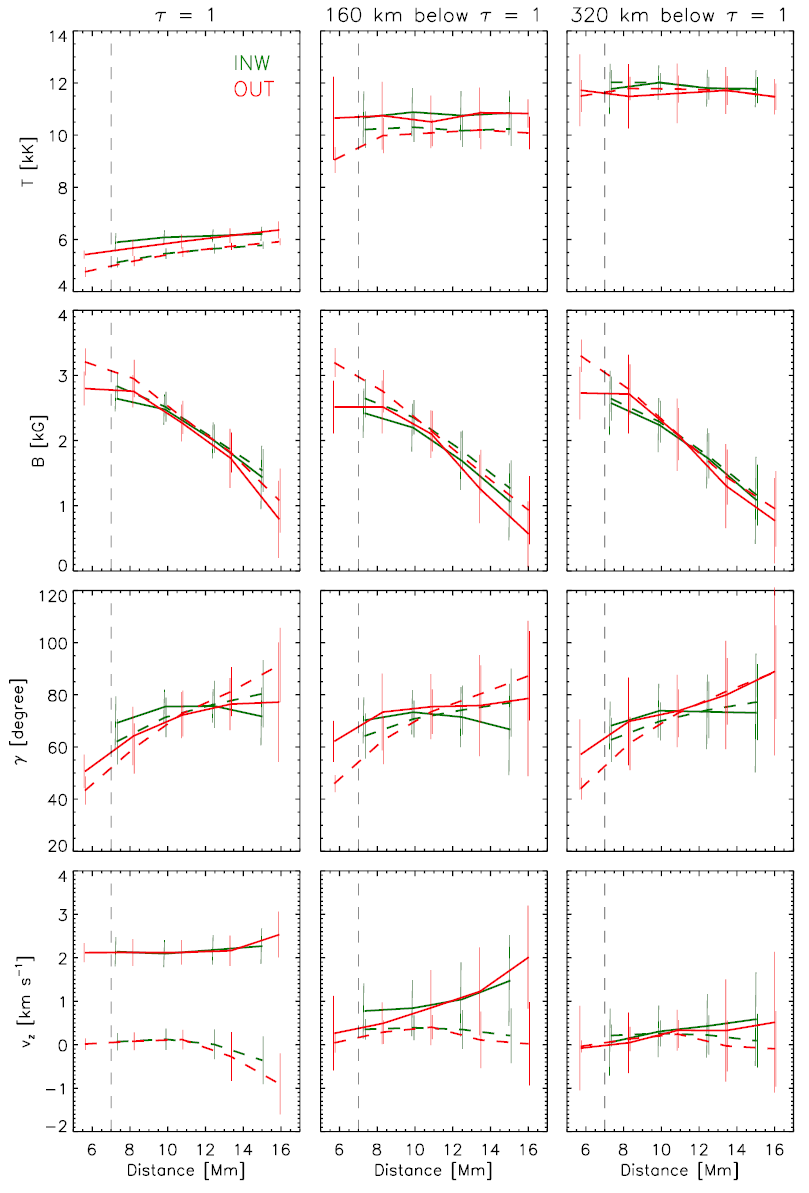}   % .eps better
\caption{Mean values and standard deviations of temperature $T$, magnetic field strength $B$, inclination $\gamma$, and vertical velocity $v_z$ versus the radial distance from the spot centre at the visible surface, $\tau_{500\rm nm} = 1$ (first column), and in two slices 160~km (second column) and 320~km (third column) below it. The values were averaged in 2.5~Mm wide sections of the radial distance. Solid lines represent the values in PGs, dashed in their surroundings, and green and red colours correspond to INW and OUT PGs, respectively. The vertical dashed lines mark the umbra/penumbra border usual for observed circular sunspots.
\label{fig:btvz_deep}}
\end{figure*}

To show changes of temperature, magnetic field magnitude, magnetic inclination, and vertical velocity in PGs and their surroundings with the radial distance from the spot centre, we calculate average values and standard deviations on the visible surface, $\tau_{500\rm nm} = 1$, and in the areas of PGs and surroundings projected down to two slices of subphotospheric layers, 160~km and 320~km below the surface. The results are displayed in Fig.~\ref{fig:btvz_deep}.
The width of the radial distance section for averaging is set to 2.5~Mm, because the number of individual PG occurrences in the time sequences of subphotospheric slices is lower (447 INW and 184 OUT PGs) than that in the sequence of surface slices due to ten times longer time step.
The plasma $\beta$ averaged in time and area of the whole penumbra is $0.9 \pm 2.0$ on the surface, $3.3 \pm 2.7$ at 160~km, and $7.3 \pm 3.5$ at 320~km below the surface. For comparison, the average plasma $\beta$ in the granulation around the sunspot is $80 \pm 10$ at the surface, $170 \pm 10$ at 160 km, and $220 \pm 10$ at 320~km below the surface.

The temperature of PGs on the visible surface is approximately $6000 \pm 300$~K, it increases with the radial distance, and INW PGs are hotter on average by 300~K than OUT PGs in the inner penumbra. The temperature in the surroundings is lower than that of PGs by 500~K on average. The temperatures obtained in these simulations are about 400~K lower than those reported by \citet{Tiwari2013} from observations. At a depth of 160~km, the temperature of PGs is approximately $10700 \pm 1000$~K and is independent of the radial distance. The temperature in the surroundings is lower than that of PGs by 600~K. Deeper, at 320~km, the temperature of $12000 \pm 1000$~K is approximately equal for all PGs and surroundings, so the thermal signature of PGs disappears there.

The magnetic field magnitudes of PGs and their surroundings decrease with the radial distance from approximately 2800~G to 1000~G and they are similar each to other within the standard deviation of their individual scatter of 300--500~G. These values are larger than the observed ones \citep[e.g.,][]{Tiwari2013} by 700--1000~G in the inner and middle penumbra, given by the simulation setting mentioned in Sect.~\ref{sec:numer}. On the visible surface, the average field in PGs is slightly weaker by 200~G than the surrounding one in the inner and outer penumbra and equally strong in the middle penumbra. The largest average difference of 250 G in the whole penumbra is seen at a depth of 160~km and practically no difference is observed deeper at 320~km. The PGs signature of lower magnetic field magnitude is weak in general, being more pronounced at 160~km below the photosphere (plasma $\beta \simeq 3$).

The distributions of magnetic field inclinations are similar each to other from the visible surface (cf. Fig.~\ref{fig:inc_rdist}) down to 160 and 320~km below it. The sign of the differences $\gamma_{\rm PG} - \gamma_{\rm s}$ of INW PGs changes from positive to negative in the middle penumbra at the radial distances of 11~Mm (depth 160~km) and 12~Mm (depth 320~km). For the OUT PGs, the differences $\gamma_{\rm PG} - \gamma_{\rm s}$ in subphotospheric layers are larger in the inner and smaller in the outer penumbra when compared to those on the surface. Their sign changes at the distance of 11~Mm. The standard deviations characterizing the scatter of individual inclinations in deep layers are larger than those on the surface: $8^\circ$--$20^\circ$ with maxima of $30^\circ$ near the outer penumbral border.

The vertical components of velocities of PGs and their surroundings are clearly distinguished on the visible surface thanks to the condition $v_z > 1.5$\,km\,s$^{-1}$ used to define PGs. The upflow velocities of INW and OUT PGs (2.1~km~s$^{-1}$) are equal and practically constant with the radial distance. The mean vertical velocities in the surroundings are close to zero at the surface and show downflows of $-0.5$\,km\,s$^{-1}$ in the outer penumbra. At the depth of 160~km, the upflows in PGs are weaker than those at the surface and increase with the radial distance from 0.5 to 1.5\,km\,s$^{-1}$, while the surrounding vertical velocities are close to zero. At 320~km below the surface, all the vertical velocities are small, between zero and 0.4\,km\,s$^{-1}$, and the vertical-velocity signature of PGs disappears. The standard deviations of individual values are in the range 0.2--0.5\,km\,s$^{-1}$ at the surface and by a factor of two larger in deeper layers.

Now let us move from individual occurrences to the evolution of PGs at the visible surface.
Lifetimes of the selected INW and OUT PGs in the simulation are in the ranges of 3--21 min and 3--13 min, respectively. The initial class of many PGs varies during their lifetime. This happens to 39\% of INW and 45\% of OUT PGs. This effect was not revealed in observations because the cadence of spectropolarimetric measurements is too low and usually only single snapshots of inclination maps are available \citep[like in][]{Sobotka2024}. To characterise the classes of PGs during their whole lifetime, we define the time-averaged class $\bar{c} = (n-m)/(n+m)$, where $n$ and $m$ are numbers of occurrences of the classes \mbox{1 and $-1$}, respectively, during the PG's lifetime. The time-averaged classes take values $-1 \leq \bar{c} \leq 1$.

\begin{figure}[t]\centering
\includegraphics[width=0.49\textwidth]{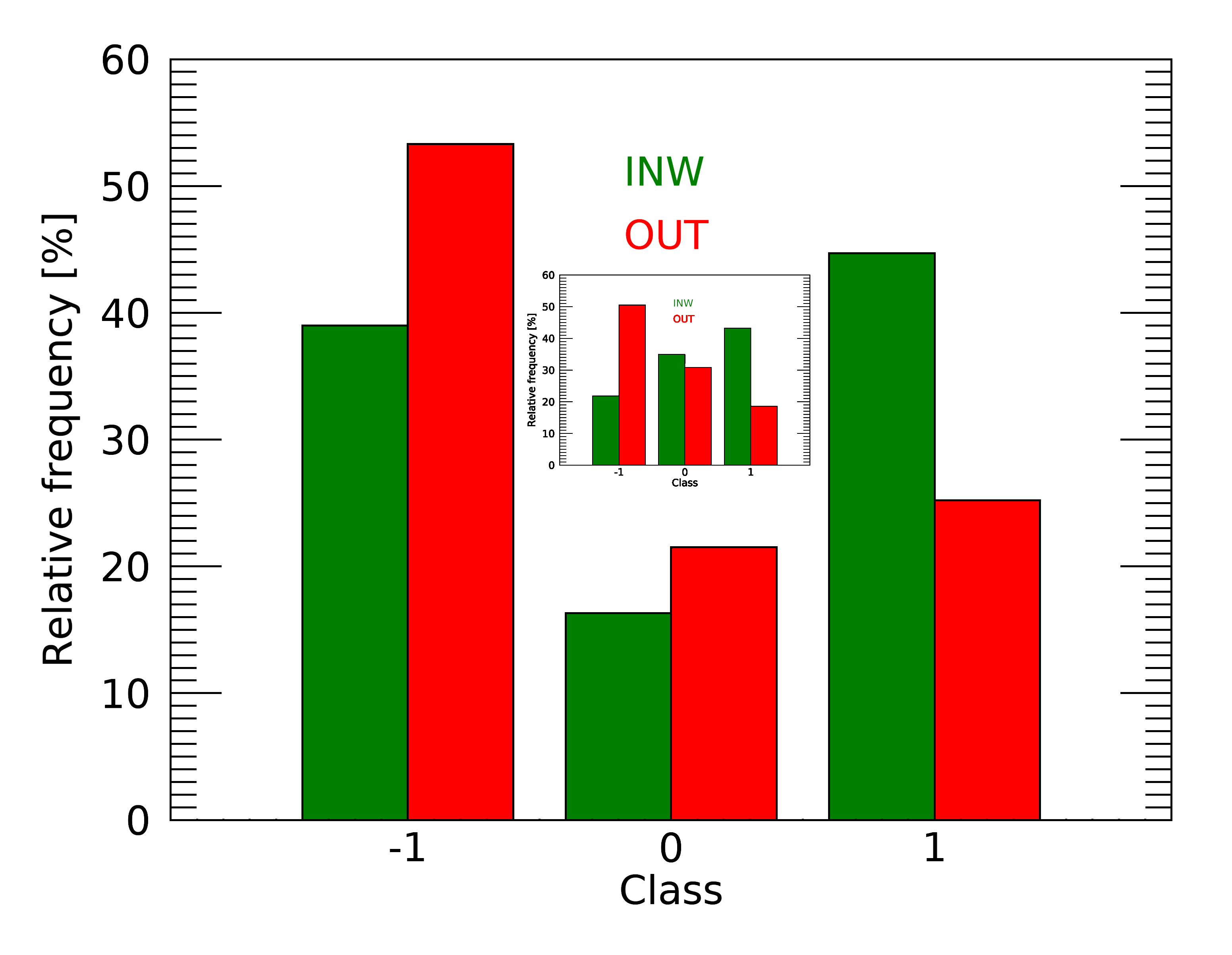}
\caption{Comparison of the magnetic field inclinations $\gamma_{\rm PG}$ and $\gamma_{\rm s}$ of 226~inward-moving (green) and 107~outward-moving (red) PGs at $\tau_{500\rm nm} = 1$. Class~$-1$: $\gamma_{\rm PG} < \gamma_{\rm s}$; class~0: indefinite cases; class~1: $\gamma_{\rm PG} > \gamma_{\rm s}$. The inset shows an analogous result derived from observations \citep[][ Fig.~2]{Sobotka2024}.
\label{fig:classhist}}
\end{figure}

We construct a histogram of $\bar{c}$ assigning the final class $-1$ to the range $-1 \leq \bar{c} < -0.5$, class 1 to $0.5 < \bar{c} \leq 1$, and class 0 to $-0.5 \leq \bar{c} \leq 0.5$. This means that the PGs with class $-1$ appearing in at least three-quarters of occurrences during their lifetime are considered to have $\gamma_{\rm PG} < \gamma_{\rm s}$. The PGs with class 1 appearing in at least three-quarters of occurrences are considered to have $\gamma_{\rm PG} > \gamma_{\rm s}$. The PGs with class 0 change their classes in such a way that we cannot determine the resulting differences between the magnetic inclinations $\gamma_{\rm PG}$ and $\gamma_{\rm s}$. The histogram is displayed in Fig.~\ref{fig:classhist}. Its form is similar to Fig.~2 of \citet{Sobotka2024} shown in the inset, which was derived from observations.
Of 226 INW PGs, 44.7\% have magnetic inclinations larger than their surroundings, 39.0\% smaller, and 16.3\% variable. Of 107 OUT PGs, 53.3\% have inclinations smaller than their surroundings, 25.2\% larger, and 21.5\% variable. This result is consistent with that obtained from the observations by \citet{Sobotka2024}.

\begin{figure}[t]\centering
\includegraphics[width=0.49\textwidth]{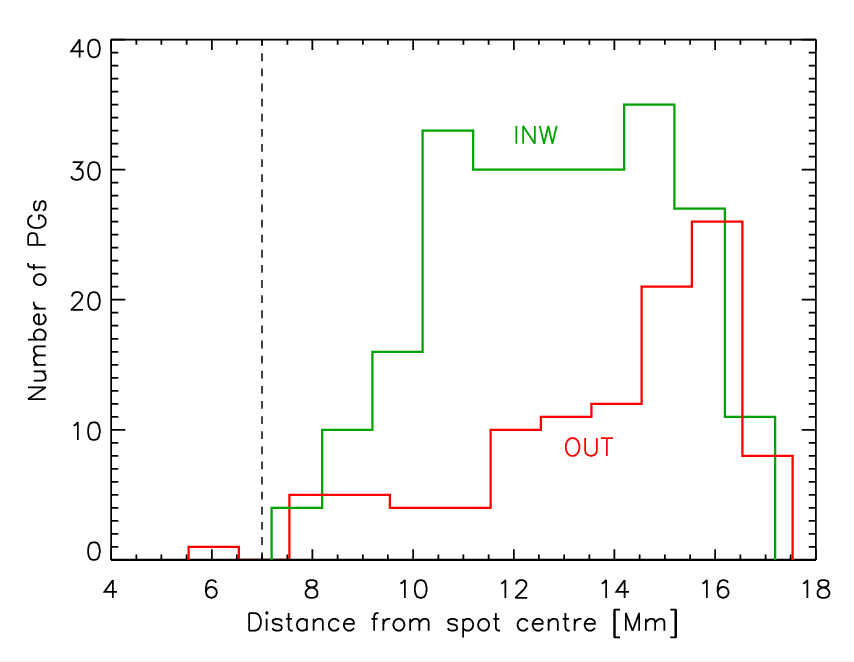}
\caption{Histogram of populations of 226 inward (green) and 107 outward-moving (red) PGs at different radial distances from the sunspot centre. The vertical dashed line shows the position of the umbra/penumbra boundary usual for observed circular sunspots ($0.4\, r_{\rm spot}$).
\label{fig:rdist}}
\end{figure}

The dependence of time averaged magnetic inclinations versus radial distance from the spot centre is practically identical to that shown in Fig.~\ref{fig:inc_rdist}.
The distribution of radial distances of PGs from the sunspot centre is illustrated in a histogram in Fig.~\ref{fig:rdist}. The vertical dashed line marks 40\% of the spot radius, which corresponds in observations of circular sunspots to the usual location of the boundary between the umbra and penumbra \citep[][p. 40]{ThoWe08}. We immediately see the lack of simulated PGs in the inner penumbra. The INW PGs appear mostly in the middle and outer penumbra and their distribution is approximately symmetric and flat between the distances 10 and 15~Mm. The OUT PGs are mostly concentrated in the outer penumbra, but they also appear in the middle and some in the inner penumbra. Compared to real sunspots, many INW PGs with $\gamma_{\rm PG} > \gamma_{\rm s}$ are missing in the simulated inner penumbra, which distorts the statistics of inclination differences for the INW PGs and explains why the frequencies of INW classes -1 and 1 are close each to other (Fig.~\ref{fig:classhist}).

The speeds of INW and OUT PGs' apparent horizontal motions are scattered around the mean values of $-1.3 \pm 0.7$ km\,s$^{-1}$ and $1.2 \pm 0.7$ km\,s$^{-1}$, respectively, and do not show any dependence on the radial distance from the spot centre.
The mean travel distances of INW and OUT PGs are 370~km and 310~km and maxima 1280~km and 710~km, respectively.

%%%%%%%%%%%%%%%%%%%%%%%%%%%%%%%

\subsection{Individual cases}
\label{sec:indiv}

\begin{figure}
  \includegraphics[width=0.49\textwidth]{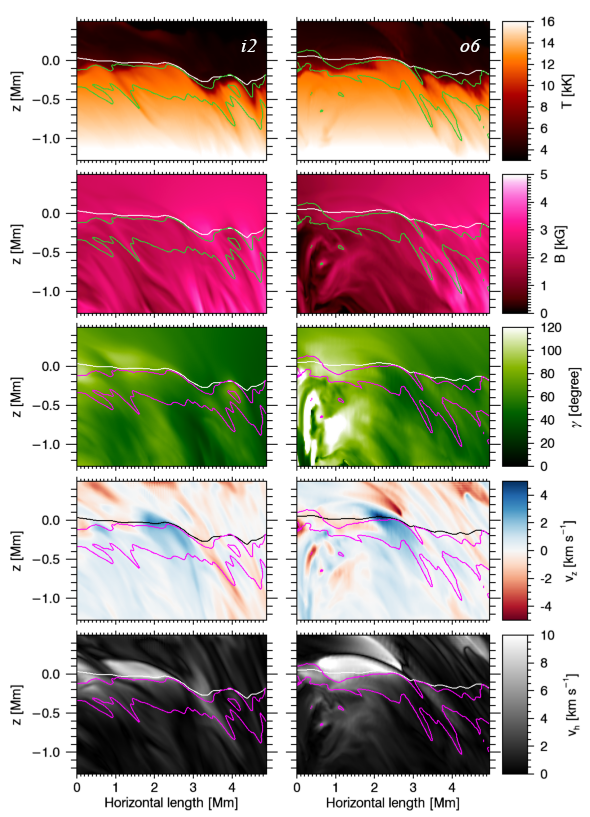}
\caption{Two examples of static vertical cuts across PGs {\it i2} (left) and {\it o6} (right) in temperature $T$, magnetic field strength $B$, magnetic inclination $\gamma$, vertical velocity component $v_{\rm z}$, and horizontal velocity component $v_{\rm h}$. Solid white or black lines correspond to $\tau_{500\rm nm}=1$ and green or magenta lines to contours of plasma $\beta = 1$ and 5. Upward-directed vertical velocities are positive (blue). Figures of all 15 vertical cuts are available {\it online}. % Fig_vcuts.zip
\label{fig:static}}
\end{figure}

To study vertical stratifications of the temperature, magnetic field strength, magnetic inclination, and vertical and horizontal components of velocity in and around PGs we made 15 static vertical plane cuts through the snapshot of the simulation box. The positions of the cuts are shown in Fig.~\ref{fig:cutmap}. Examples of two cuts {\it i2} and {\it o6} are displayed in the left and right columns of Fig.~\ref{fig:static}, respectively. The solid white or black lines depict the visible surface ($\tau_{500\rm nm}=1$) and green or magenta lines the contours of plasma $\beta = 1$ and 5 in all panels. Positive vertical velocities (blue) are directed upwards. The axis of horizontal length along the cut is always oriented towards the umbra, which means, that in the cuts, the umbra is on the right. Figures of all 15 vertical cuts are available {\it online}. % Fig_vcuts.zip

%----------------------------------------------------
\begin{table}\centering
\caption{Properties of individual penumbral grains in vertical cuts.}  \label{tab:pgstat}
%\begin{tabular}{lclccccc} - reduced left-most space from 6pt to 4pt, removed right-most space
\begin{tabular}{@{\hspace*{4pt}}lclccccc@{}}
\hline\hline
PG  & Pos. &  Penumbra & \multicolumn{3}{c}{Inclination} & \multicolumn{2}{c}{Flows} \\
    & [Mm] &           &      deep & surface & high & (1) & (2) \\
\hline
\multicolumn{8}{l}{Inward-moving PGs} \\
  i1  & 3.0 &     mid   &      $-$ &  $+$   &  $+$   &  y  &  nn \\
  i2  & 2.5 &    inner  &      $-$ &  $+$   &  $+$   &  n  &  y \\
  i3  & 2.3 &     mid   &      $-$ &  $+$   &  $+$   &  y  &  y \\
  i4  & 3.0 &     mid   &      $-$ & $\sim$ & $\sim$ &  y  &  n \\
  i5  & 3.2 &     mid   &      $-$ &  $+$   &  $+$   &  y  &  y \\
  i6  & 2.6 &     mid   &      $-$ &  $+$   & $\sim$ &  y  &  y \\
  i7  & 3.0 &    inner  &      $-$ &  $+$   & $\sim$ &  n  &  y \\
  i8  & 4.8 &   inner-mid &    $-$ &  $+$   &  $+$   &  y  &  y \\
  i9  & 2.5 &     mid   &      $-$ &  $+$   &  $+$   &  n  &  nn \\
\hline
\multicolumn{8}{l}{Outward-moving PGs} \\
  o1  & 3.5 &     mid   &      $-$ & $\sim$ &  $+$   &  y  &  y \\
  o2  & 3.0 &    outer  &      $-$ &  $-$   & $\sim$ &  y  &  n \\
  o3  & 2.5 &    inner  &      $+$ &  $+$   &  $+$   &  n  &  y \\
  o4a & 3.2 &     mid   &      $+$ & $\sim$ & $\sim$ &  n  &  nn \\
  o4b & 2.0 &    outer  &      $-$ & $\sim$ &  $+$   &  y  &  y \\
  o5a & 3.5 &    inner  &      $-$ &  $-$   &  $-$   &  n  &  nn \\
  o5b & 2.0 &    outer  &      $-$ &  $-$   &  $+$   &  y  &  y \\
  o6  & 2.5 &     mid   &      $-$ &  $-$   &  $+$   &  y  &  y \\
\hline
\end{tabular}
\tablefoot{Pos. -- an approximate horizontal position of a PG along the cut. Magnetic inclination in a PG below, at, and above the visible surface compared to the surroundings: enhanced ($+$), similar ($\sim$), reduced ($-$). Properties of gas flows: (1) -- the upflow is co-spatial with a reduced inclination below the surface, yes/no; (2) -- the Evershed flow is co-spatial with an enhanced inclination at and above the surface, yes/no; {\em nn} means that the Evershed flow is too weak or absent.}
\end{table}
%-----------------------------------------------------

The cut {\it i2} shows an INW PG (horizontal position 2.5~Mm) in the inner penumbra. In deep layers ($z < -400$~km), its magnetic inclination is locally reduced and around the surface enhanced.
There is an upflow in the PG region. A large horizontal velocity, which depicts the Evershed flow, appears at and above the surface being co-spatial with the enhanced magnetic inclination. Another small temperature increase (horizontal position 4~Mm), probably a peripheral umbral dot, shows a weak upflow and enhanced inclination. The cut {\it o6} includes a bright OUT PG (horizontal position 2.5 Mm) located in the middle penumbra. Its magnetic inclination is reduced at and below the visible surface, being co-spatial with an upflow. The enhanced inclination above the surface (position 0.5--2.3~Mm) is spatially correlated with a region of large horizontal velocity (the Evershed flow).

The relevant physical characteristics of nine INW and eight OUT PGs (we note that the cuts {\it o4} and {\it o5} intersect two OUT PGs each) are summarised in Table~\ref{tab:pgstat}. We show the location of PGs in the penumbra (inner, middle, outer) and give the local enhancements (+) and reductions ($-$) of the PG's magnetic inclination
in subphotospheric layers, at the visible surface, and higher in the photosphere. Furthermore, we look for the properties of gas flows in and around PGs, namely, (1) whether the upflow in PGs is co-spatial with the reduced inclination below the surface, and (2) whether the Evershed flow is co-spatial with the enhanced inclination at and above the surface.
The following findings can be derived from the 15 {\it online} % Fig_vcuts.zip
figures of the cuts and Table~\ref{tab:pgstat}:

\begin{figure*}[t]\centering
  \includegraphics[width=0.98\textwidth]{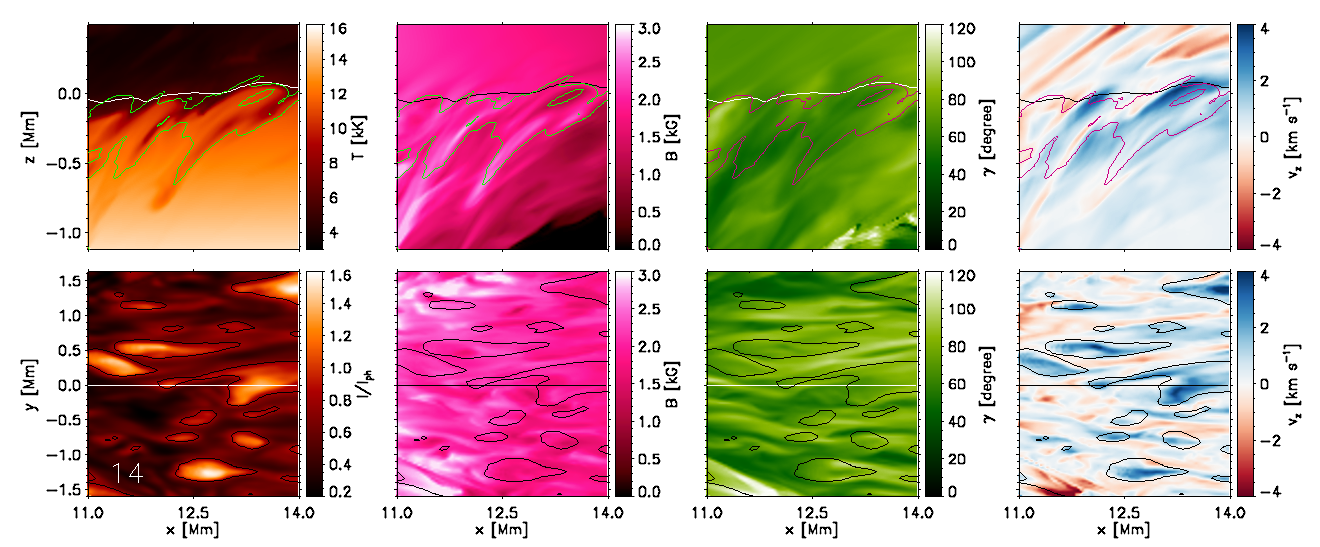}
\caption{Snapshot of vertical $xz$-plane movie {\it mv07}, frame 14. Top row: vertical cuts of PGs 7b and 7a in temperature $T$, magnetic field strength $B$, magnetic inclination $\gamma$, and vertical velocity component $v_{\rm z}$. Solid white or black lines correspond to $\tau_{500\rm nm}=1$ and green or magenta lines to contours of plasma $\beta = 1$ and 5. Upward-directed vertical velocities are positive (blue). The horizontal distance $x$ has the origin in the spot centre. Bottom row: corresponding horizontal slices in normalised intensity, magnetic inclination, and vertical velocity at $\tau_{500\rm nm} = 1$. Contours correspond to $I_{\rm out} = 0.75\, I_{\rm ph}$ and a horizontal line marks the position of the $xz$ cut. Ten movies of 11 PGs are available {\it online}. % vmv01-10.zip
\label{fig:movie}}
\end{figure*}

(i) In many cases (5 INW PGs of 9, 7 OUT PGs of 8), upflows appear along the whole length of the bright part of the penumbral filament, not only in a PG. 

(ii) In the subsurface layers, the magnetic inclination below PGs is locally reduced (9 INW and 6 OUT). This region of reduced inclination is often co-spatial with upflows (6 INW, 5 OUT). 

(iii) At the visible surface, the inclination of INW PGs is locally enhanced (8 cases of 9) and the inclination of OUT PGs is reduced or unchanged (7 cases of 8). This relation between magnetic inclination and direction of PGs' motion (Sect.~\ref{sec:stat}) is valid only in the surface layer (cf. Table~\ref{tab:pgstat}). 

(iv) Above the surface, in the photosphere, the inclination along penumbral filaments is larger than or equal to the surrounding one (9 INW, 7 OUT), and the regions of enhanced inclination are often connected with the Evershed flow (6 INW, 5 OUT). 

(v) The magnetic field strength below PGs is locally reduced in more than half of the cases (5~INW, 5~OUT). 

(vi)~The plasma $\beta = 1$ level is close to the $\tau_{500\rm nm}=1$ surface and the Evershed flow is formed only in regions with $\beta < 1$.

The time sequence of 101 vertical cuts in the $xz$ plane crossing the sunspot centre allows us to make movies of vertical cuts through PGs with the cadence of 18~s. We made 10 movies available {\it online}
% vmv01-10.zip
with 11 PGs (two in the movie {\it mv07}) intersected by the cuts. The panels in the top row of the movies show vertical cuts in temperature, magnetic field strength, magnetic inclination, and vertical velocity component. Solid white or black lines depict the visible surface \mbox{($\tau_{500\rm nm}=1$)} and green or magenta lines correspond to contours of plasma $\beta = 1$ and 5. The bottom row displays corresponding horizontal slices in intensity, magnetic field strength, magnetic inclination, and vertical velocity at $\tau_{500\rm nm} = 1$ together with a line marking the position of the $xz$ cut. The contour corresponds to $I_{\rm out} = 0.75\, I_{\rm ph}$. The horizontal coordinate $x$ gives the distance from the sunspot centre and when it is positive, the umbra is on the left side and vice versa. A snapshot (frame~14) of the movie {\it mv07} is displayed in Fig.~\ref{fig:movie}. The cut intersects the INW PG\,7b (left) and OUT PG\,7a (right). 

We can see in the movies that the apparent motions of PGs are related to horizontal displacements of columns with ascending hot plasma in subphotospheric layers, where plasma $\beta$ is in the range 1--5, except for regions with strong turbulence. These values of plasma $\beta$, where gas and magnetic pressures are of the same order, enable the magnetic field to modify the plasma motions.

%----------------------------------------------------
\begin{table}\centering
\caption{Properties of penumbral grains in vertical plane movies.}  \label{tab:pgmov}
\begin{tabular}{lccrcc}
\hline\hline
Movie &  Frame   &   PG  &  $x$ [Mm] &  Motion   &  Class           \\
\hline
mv01  &   5--20  &   1   &  $-14.0$  &    OUT    &  $-1$       \\
mv02  &  66--100 &   2   &  $-10.0$  &    INW    &    1        \\
mv03  &  12--48  &   3   &  $-11.0$  & INW$\rightarrow$stop & 1$\rightarrow-1$ \\ 
mv04  &  78--95  &   4   &  $ -8.8$  &    INW    &    1        \\
mv05  &  17--34  &   5   &    9.0    &    OUT    &    1        \\
mv06  &  33--43  &   6   &   10.0    &    OUT    &    0        \\
mv07  &   0--17  &   7a  &   13.0    &    OUT    &  $-1$       \\
      &          &   7b  &   12.3    &    INW    &   1         \\
mv08  &  56--83  &   8   &   13.5    &  INW$\rightarrow$OUT & 1$\rightarrow-1$  \\
mv09  &  74--91  &   9   &   15.0    &    INW    &  $-1$       \\
mv10  &  65--74  &  10   &   10.5    &    INW    &   1         \\
\hline
\end{tabular}
\tablefoot{$x$ -- distance from the spot centre; class~$-1$: $\gamma_{\rm PG} < \gamma_{\rm s}$; class~0: indefinite; class~1: $\gamma_{\rm PG} > \gamma_{\rm s}$.}
\end{table}
%----------------------------------------------------

The occurrences of PGs in individual cuts (frames) of the whole sequence, their positions $x$, directions of motion, and classes (Sect.~\ref{sec:stat}) are listed in Table~\ref{tab:pgmov}. The classes were estimated visually from the horizontal slices of inclination and corresponding vertical cuts around the visible surface.
The PGs are located in the inner ($x < 10$\,Mm), middle ($10 < x < 13$\,Mm), and outer ($x > 13$\,Mm) penumbra.
Two PGs, 1 and 9, are close to the penumbra-granulation boundary. We note that the PG\,4 is located 100~km off the vertical cutting plane. The inspection of the {\it online}  % vmv01-10.zip
movies confirms that the regions of upflows are often connected with the reduced magnetic inclination (PGs 1, 2, 3, 4, 5, 7a, 9, 10). The magnetic field strength near the surface is locally reduced in PGs or weakens with time (in PGs\,2 and 6). It is also confirmed that the class can change in time (Sect.~\ref{sec:stat}). This effect is observed in the PGs\,3 and 8, which were initially assigned class~1 and then changed to class~$-1$. We learn from the movies that such a change of class can be accompanied by a change of the direction of apparent motion. Indeed, PG\,3 initially moved inwards and finally stopped. Similarly, PG\,8 moved inwards, stopped, and then moved outwards.

\begin{figure}
  \includegraphics[width=0.49\textwidth]{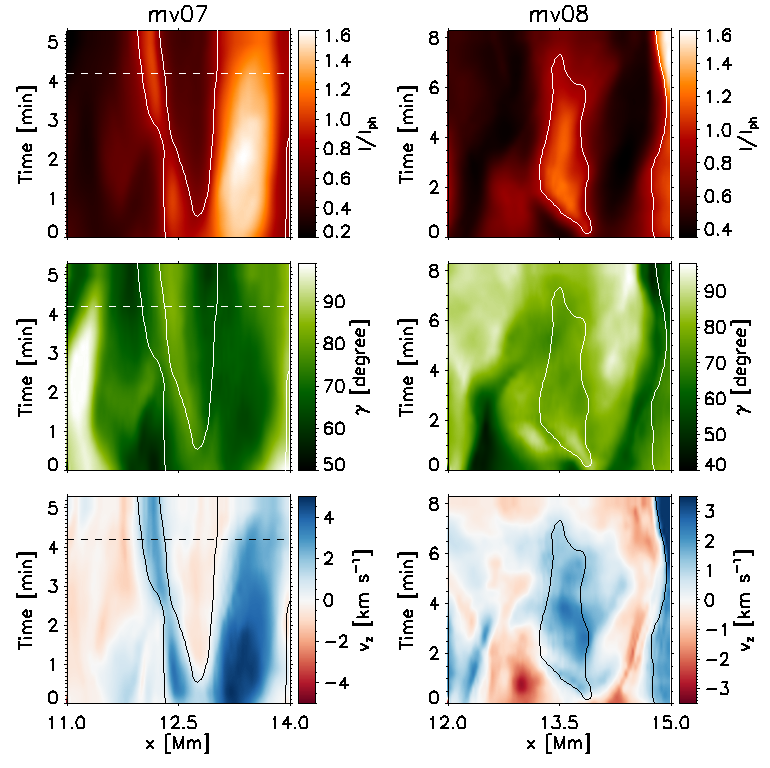}
\caption{Time slices of movies {\it mv07} (left column) and {\it mv08} (right colummn) showing the apparent motions of structures seen in the normalised intensity $I_{\rm out}/I_{\rm ph}$, magnetic inclination $\gamma$, and vertical velocity $v_{\rm z}$ at $\tau_{500\rm nm} = 1$. The time slice of mv07 shows the inward motion of PG\,7b (left) and the outward motion of PG\,7a (right). Contours correspond to $I_{\rm out} = 0.75\, I_{\rm ph}$ and horizontal dashed lines mark the time corresponding to frame 14 of {\it mv07} displayed in Fig~\ref{fig:movie}. The intensity slice of {\it mv08} shows the changing direction of PG\,8. Contours correspond to $I_{\rm out} = 0.9\, I_{\rm ph}$.
\label{fig:tslice}}
\end{figure}

Time slices displayed in Fig.~\ref{fig:tslice} demonstrate apparent horizontal motions of intensity, magnetic inclination, and vertical velocity structures of PGs 7a, 7b, and 8 on the visible surface. Tracks of enhanced intensity, magnetic inclination, and upflow corresponding to INW~PG\,7b are clearly seen in the left column ({\it mv07}, position $x = 12$--12.5~Mm). The neighbouring OUT~PG\,7b ({\it mv07}, $x = 13$--13.5~Mm) is characterised by enhanced intensity and upflow but reduced magnetic inclination. The change of direction of the apparent motion of PG\,8 ({\it mv08}, right column) can be seen in the intensity track. During the first two minutes, PG\,8  moved inwards, then stopped for one minute and moved slowly outwards. Traces of the fast inward motion in the first two minutes can also be observed in the figure as small paths of enhanced magnetic inclination and upflow.

Apart from the two changeable PGs\,3 and 8, there are four INW PGs of class~1 and one (PG\,9) of class~$-1$. Two of four OUT PGs are of class~$-1$, one (PG\,6) of class~0, and one (PG\,5) of class~1. This is consistent with the statistical result shown in Fig.~\ref{fig:classhist} and with the result of observations. The physical conditions around the non-matching PGs are specific: proximity to the penumbra-granulation boundary (PG\,9) or to a large region of strongly reduced inclination (PGs~5 and 6). Moreover, the latter PG is located in a turbulent environment. The turbulent character of subsurface motions is seen in the movies, particularly in the outer and middle penumbra.
This explains the large scatter of individual values of $\gamma$ and $v_{\rm z}$ near the outer penumbral border (Sect.~\ref{sec:stat}).
The turbulence affects shapes and displacements of the subphotospheric columns of rising hot plasma (PGs 2, 3, 6, 8, 9, 10) and this is another factor that can influence the apparent motions of PGs.

%%%%%%%%%%%%%%%%%%%%%%%%%%%%%%%%%%%%%%%%%%%%%%%%%%%%%%%%%%%%%%%%%%%%%%%%

\section{Discussion and conclusions}
\label{sec:discuss}

Using the results of numerical MHD simulations of sunspots, we tried to describe physical conditions in and around PGs, including those that may influence the direction of apparent motions of penumbral grains. The results of spectropolarimetric observations presented by \citet{Sobotka2024} indicate that the difference between magnetic inclinations of PGs and their surroundings is an important factor: most of the inward moving PGs have an inclination larger than that in their surroundings and most of the outward moving PGs have an inclination smaller than the surrounding one. However, \citet{Sobotka2024} reported that this relation failed in about 20\% of PGs, and there was also a large number of cases when the difference of inclinations was unclear. The simulations can provide a different perspective, because they describe physical conditions with such a high spatial and temporal resolution that observations cannot reach.

We must rely on the assumption that the simulations are sufficiently realistic, meaning that the general differences between simulated and observed sunspots have little influence on our observables. 
Throughout the penumbra, the magnetic fields in the simulation are stronger on average than in observed sunspots. 
The increase of the azimuthally averaged field inclination with the radial fraction is consistent between simulations and observations. 
However, the radially averaged brightness of our simulation increases slower than in observations, meaning that at relative radii, where in observations the spot has the brightness of the inner penumbra and its vigorous convection, the simulation is significantly darker and less convectively active.
Therefore, fewer PGs occur in simulations at places with inclinations that in observed spots correspond to the inner penumbra. Hence, we are missing many class~1 INW PGs in the inner penumbra. The tightly packed inward-moving PGs observed at the umbra-penumbra border in real spots are missing, only a few solitary PGs are present there, and the simulated umbral border is strongly fragmented, quantified as the fractal dimension in Table~2 of \citet{2020A&A...638A..28J}. %Jurčák
This distorts the statistics of inclination differences for the INW PGs (Sect.~\ref{sec:stat}).
The closer we approach the outer edge of the penumbra, the less the difference in average inclination between observation and simulation becomes. Given that the vertical magnetic component, $B_\mathrm{ver}$, is too low to inhibit the penumbral type of convection there \citep[see e.g.][]{2018A&A...611L...4J,2018A&A...620A.104S}, % Jurčák, Schmassmann
the excess field strength has little influence on the type of convection, including the INW and OUT PGs. 

%------------------------------

The statistical study of inclinations in individual occurrences of PGs and surrounding areas shows a large scatter of values that indicates a dynamic character of the convection. Mean values of inclinations are obtained by averaging in time and over sections of radial distance from the spot centre. The mean values in PGs change weakly with the radial distance, while the mean values in surrounding areas clearly increase with the radial distance (Fig.~\ref{fig:inc_rdist}). The average differences of inclinations, $\gamma_{\rm PG} - \gamma_{\rm s}$, between PGs and their surroundings at the visible surface are quite small and comparable with standard deviations of the scatter of individual values. According to the bottom panel of Fig.~\ref{fig:inc_rdist}, the signs (classes) of $\gamma_{\rm PG} - \gamma_{\rm s}$ depend generally on the radial distance, being positive in the inner and negative in the outer penumbra. In the middle penumbra, INW PGs change their sign from positive to negative at a larger distance from the umbra than OUT PGs. Similar characteristics of average inclinations are found also in deeper layers, 160~km and 320~km below $\tau_{500\rm nm} = 1$.
Under these circumstances, the result of the statistics shown in Fig.~\ref{fig:classhist} may arise from the change of classes along the radial distance combined with different populations of INW and OUT PGs at different distances (Fig.~\ref{fig:rdist}) and the simulations may not prove unambiguously the relation between the difference of inclinations $\gamma_{\rm PG} - \gamma_{\rm s}$ and the direction of apparent motion.

From the detailed study of individual PGs in the simulations
we newly find that the difference of inclinations (the class) may vary during the lifetime of a PG and even the direction of the PG's motion can change from inwards to outwards (we cannot rule out changes in the opposite direction either). This may explain the indefinite cases (class 0) in the simulations and also in the results of observations -- however, the observational limits and the accuracy of inversions are important, too. Random disturbances are also introduced by the turbulence, which is present in the outer and middle penumbra, and is strongest in the layers around the visible surface. It significantly influences the dynamics of the hot plasma upflows and the magnetic field vector, as seen in most of the {\it online} videos. % vmv01-10.zip

Based on the results of numerical simulations, we find a certain difference between the INW and OUT PGs at the visible surface, namely, that the average difference $\gamma_{\rm PG} - \gamma_{\rm s}$ of INW PGs remains positive (class 1) over a greater distance from the umbra than in the case of OUT PGs (Fig.~\ref{fig:inc_rdist}, bottom panel). However, due to a large scatter of individual inclinations, we can neither confirm nor deny that the difference of magnetic inclinations and the direction of apparent motion are directly related. We find that the turbulent character of magnetoconvective motions in the surface layers is an important factor that influences the apparent motions of PGs in the simulations, introducing a certain degree of randomness.

%%%%%%%%%%%%%%%%%%%%%%%%%%%%%%%%%%%%%%%%%%%%%%%%%%%%%%%%%%%%%%%%%%%%%%%%%

\begin{acknowledgements}

We thank N. Bello Gonz\'alez and the anonymous referee for valuable comments, which helped us to improve the paper. This work was supported by the Czech-German common grant, funded by the Czech Science Foundation under the project 23-07633K and by the Deutsche Forschungsgemeinschaft under the project BE 5771/3-1 (eBer-23-13412), and the institutional support ASU:67985815 of the Czech Academy of Sciences.
The simulation leading to the results obtained was run on Piz Daint at the Swiss National Supercomputing Centre, Switzerland, financed through the ACCESS program of the SOLARNET project that has received funding from the European Union Horizon 2020 research and innovation program under grant agreement no 824135.

\end{acknowledgements}

%%%%%%%%%%%%%%%%%%%%%%%%%%%%%%%%%%%%%%%%%

\bibliographystyle{aa}
\bibliography{bibliography2}

%%%%%%%%%%%%%%%%%%%%%%%%%%%%%%%%%%%%%%%%%

\begin{appendix}
\section{Initial magnetic field}
\label{sec:app_ini_mag}
The initial magnetic field uses the self-similar approach used in 
\cite{1958IAUS....6..263S} % Schlüter & Temesváry
and others:
\begin{align}
    B_z(r,z)&=B_0f(\xi)g(z),\\
    B_R(r,z)&=-B_0\frac{r}{2}f(\xi)g'(z),\\
    \xi&=r\sqrt{g(z)},\quad\text{with}\\
    f(\xi)&=\exp\left(-(\xi/R_0)^4\right),\\
    g(z)&=\exp\left(-(z/z_0)^2\right),
\end{align}
whereby $B_z$ is the vertical component of the magnetic field, $B_R$ the radial component, $r$ the distance from the spot axis, and $z$ the distance from the bottom of the box. The parameters are $B_0=6.4\,$kG, the initial field strength on the spot axis at the bottom of the box, $R_0=8.2\,$Mm and $z_0=6.4185\,$Mm, which determines how far away from the axis and the bottom $B_z$ reduces by a factor $e$. These parameters lead to a sunspot with an initial flux of $F_0=1.2\times10^{22}$Mx and a field strength at the top of 2.56\,kG. A more detailed description of the initial state of the simulation is in \cite{Rempel2012}. % 2012ApJ...750...62R
The choice of $f(\xi)$ and $g(z)$ are specific to this simulation, other self-similar sunspot models use for $f(\xi)$ a Gaussian 
\citep[e.g.][]{1958IAUS....6..263S,2005A&A...441..337S} % Schlüter & Temesváry, Schüssler & Rempel
or the boxcar function
\citep[][and many analytical models]{2020ApJ...893..113P} % Panja et. al
and $g(z)$ is sometimes related to the pressure difference between the spot center and the quiet Sun 
\citep[e.g.][]{1958IAUS....6..263S}.
\end{appendix}

\end{document}